\documentclass[preprint2]{aastex}

\usepackage{natbib}

\shorttitle{3D Coronal Slow Modes}
\shortauthors{Marsh, Walsh and Plunkett}

\begin{document}

\title{3D Coronal Slow Modes: Towards 3D Seismology}

\author{M. S. Marsh, R. W. Walsh and S. Plunkett\altaffilmark{1}}
\affil{Jeremiah Horrocks Institute for Astrophysics \& Supercomputing, University of Central Lancashire, Preston, PR1
2HE, UK}
\email{mike.s.marsh@gmail.com}

\altaffiltext{1}{Naval Research Laboratory, Code 7662, Washington, DC 20375, United States}

\begin{abstract}
On 2008 January 10, the twin Solar Terrestrial Relations Observatory (STEREO) A and B spacecraft conducted a high time cadence study of the solar corona with the Extreme UltraViolet Imager (EUVI) instruments with the aim of investigating coronal dynamics. Observations of the three-dimensional propagation of waves within active region coronal loops and a measurement of the true coronal slow mode speed are obtained. Intensity oscillations with a period of $\approx$12~minutes are observed to propagate outwards from the base of a loop system, consistent with the slow magnetoacoustic mode. A novel analysis technique is applied to measure the wave phase velocity in the observations of the A and B spacecraft. These stereoscopic observations are used to infer the three-dimensional velocity vector of the wave propagation, with an inclination of ${37 \pm 6} ^{\circ}$ to the local normal and a magnitude of $132 \pm 9$ and $132 \pm 11$ km~s$^{-1}$, giving the first measurement of the true coronal longitudinal slow mode speed, and an inferred temperature of $0.84 \pm 12$~MK and $0.84 \pm 15$~MK.
\end{abstract}


\section{Introduction}\label{sect_intro}
The almost continual observations of the Solar and Heliospheric Observatory (SOHO) and Transition Region and Coronal Explorer (TRACE) satellites have made observations of the solar corona readily available. During this time a multitude of coronal wave observations have been made, which may be interpreted as solutions to the Magneto-Hydrodynamic (MHD) wave equations as described by \cite{edw83}. This recent consonance of theory and observation has allowed the promise of using coronal seismology to derive hitherto indeterminable physical parameters of the coronal atmosphere, as suggested by \cite{uch70, rob84}. These principles have been used to estimate coronal parameters such as the magnetic field strength \citep{nak01, ash02, ver04, wan07, vdor08, ofm08}, damping mechanisms \citep{ofm02}, dissipative coefficients \citep{nak99}, Alfv\'{e}n speeds \citep{arr07} and density stratification \citep{and05}.

A considerable limitation to current coronal seismology efforts is the unknown three dimensional geometry of coronal structures, since spatial measurements are confined to the two dimensional plane perpendicular to the observers line of sight. The STEREO mission consists of two spacecraft, one which orbits the Sun ahead of the Earth (STEREO A) and another that orbits behind (STEREO B). The STEREO spacecraft allow the Sun to be observed from two different vantage points, allowing stereoscopic measurements to be made, and the three dimensional nature of coronal structures to be inferred.

\begin{figure*}[t]
\centering
\includegraphics{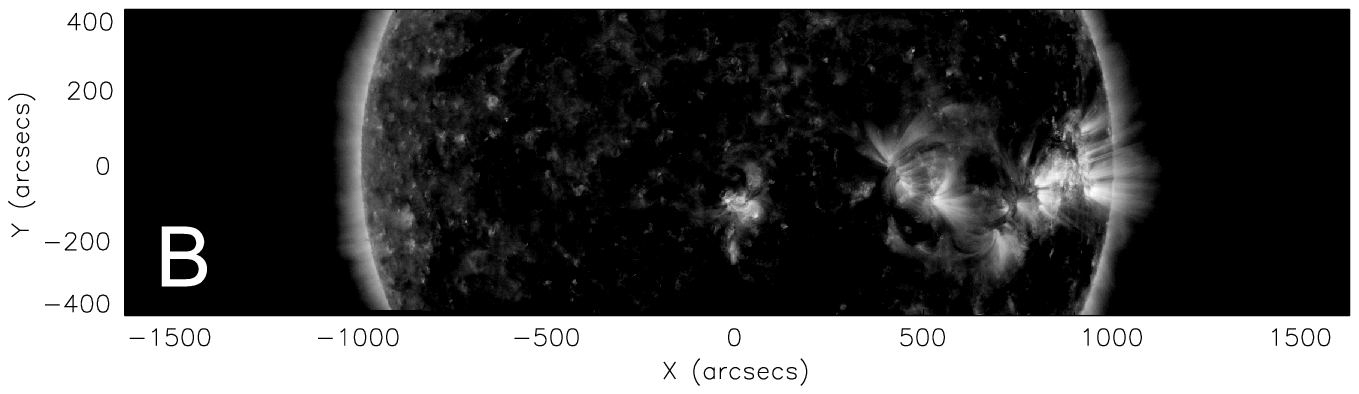}
\includegraphics{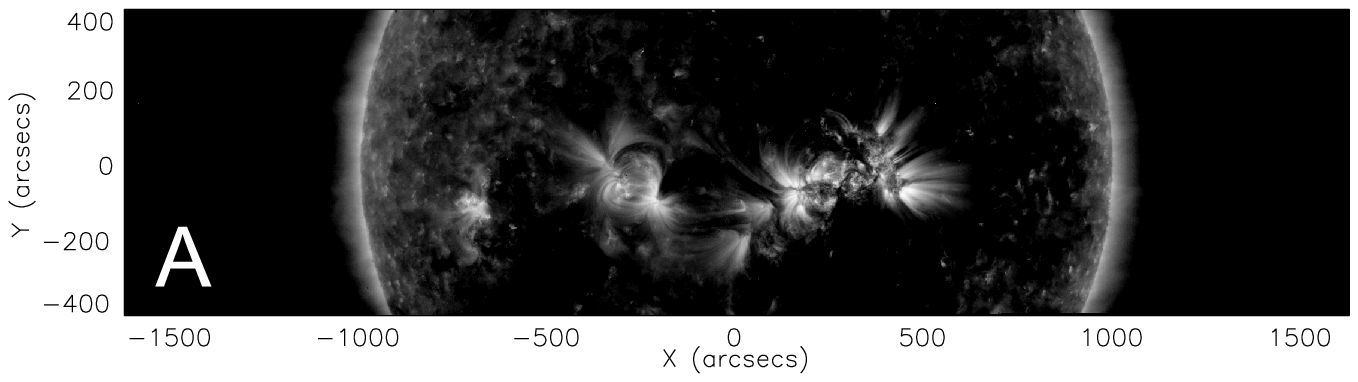}
\caption{The 2048$\times$512 pixel central strip of the Sun observed by EUVI during the high cadence observations, from the point of view of STEREO B (top) and STEREO A (bottom). The active region loops of interest are located at (390\arcsec,10\arcsec) and (-350\arcsec,10\arcsec) in heliocentric-cartesian coordinates relative to B and A respectively.}
\label{fig1}
\end{figure*}

Previous imaging observations of propagating slow-magnetoacoustic waves in coronal loops have been made using TRACE and SOHO/EIT. These are observed as intensity propagations in the coronal passbands with amplitudes on the order of a few percent and apparent propagation velocities in the range 70--235 km~s$^{-1}$ \citep[see][]{nig99, ber99, rob01, dem02a}. However, due to the inclination of the loop systems, these observed velocities are projections of the absolute speed perpendicular to the line of sight.

We present the first three dimensional observations of coronal slow-magnetoacoustic wave propagation. These observations allow us to infer the three dimensional geometry of the wave propagation vector and its magnitude, allowing the measurement of the true longitudinal slow mode speed and temperature within a coronal loop.

\begin{figure*}[t]
\centering
\epsscale{2.0}
\plottwo{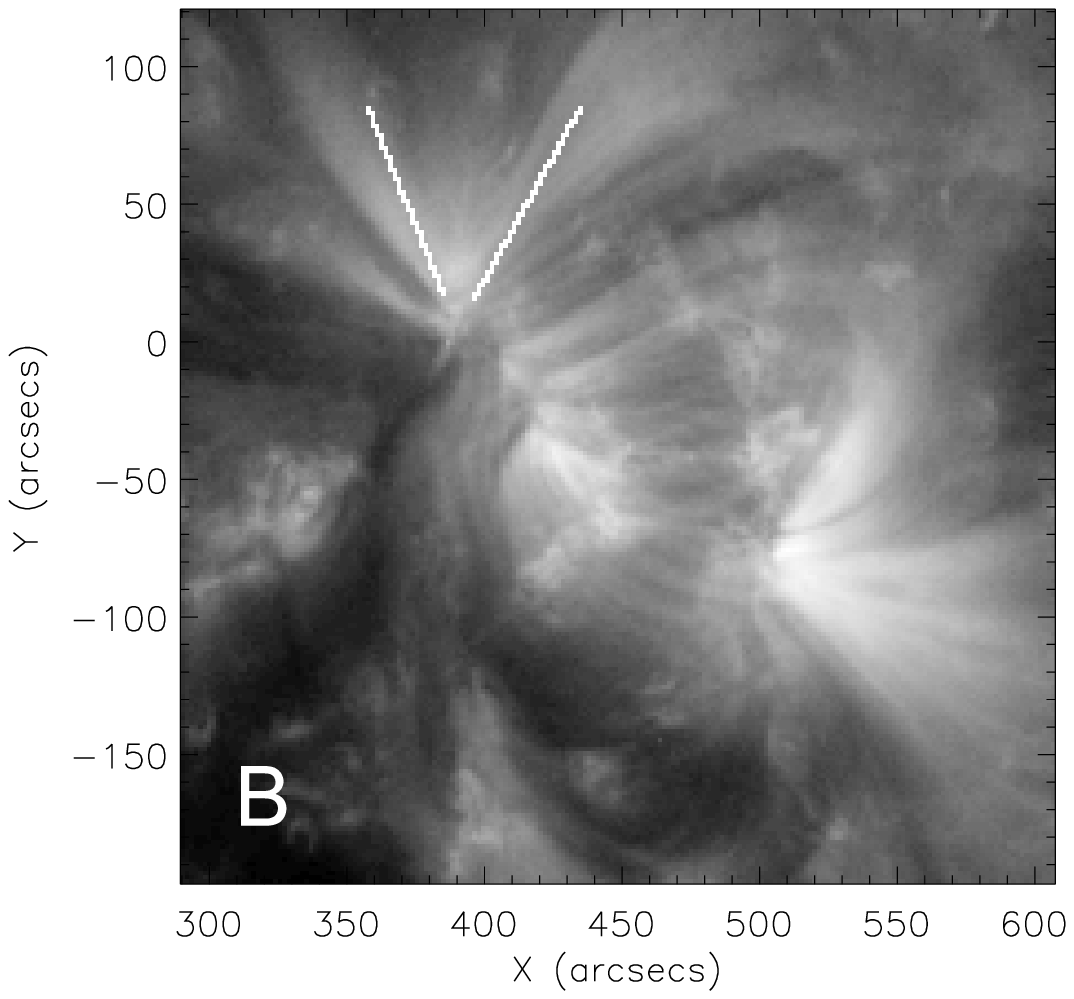}{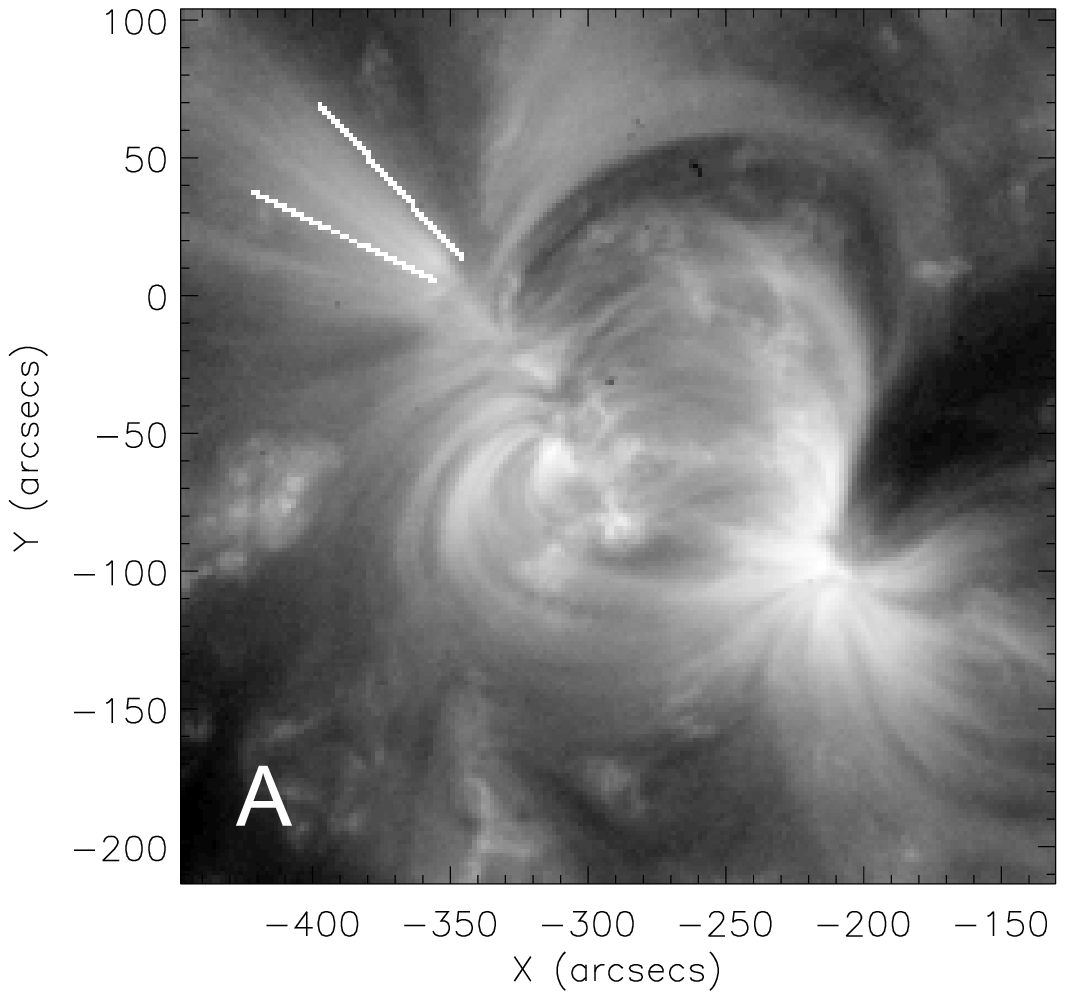}
\caption{The 200$\times$200 pixel sub-field views of the coronal loop system from STEREO B (left) and STEREO A (right). The white lines outline the same coronal loop system, observed from differing points of view.}
\label{fig2}
\end{figure*}

\begin{deluxetable}{lcc}
\tabletypesize{\scriptsize}
\tablecaption{Observational parameters of the EUVI high cadence observing program\label{tab1}}
\tablewidth{0pt}
\tablehead{
\colhead{Observational Parameter} & \colhead{STEREO A} & \colhead{STEREO B}
}
\startdata
Observations start & 2008-01-10T12:00:30.004 & 2008-01-10T12:00:50.102 \\
FOV (pixels) & $2048\times512$ & $2048\times512$ \\
Pixel scale (arcsec) & 1.588 & 1.590 \\
Distance from Sun center (m) & 1.4471540e+11 & 1.5073979e+11 \\
Image cadence (s) & 30.0 & 30.0 \\
Exposure time (s) & 4.0 & 4.0 \\
Passband (\mbox{\AA}) & $171$ & $171$ \\
Heliographic Longitude (deg)\tablenotemark{a} & 21.295628 & -22.987068 \\
Heliographic Latitude (deg)\tablenotemark{a} & -6.0627318 &  -1.0544470\\
 \tableline
\enddata
\tablenotetext{a}{Heliographic longitude and latitude are Heliocentric Earth Equatorial (HEEQ), equivalent to the Stoneyhurst heliographic coordinates.}
\tablecomments{During the observations the spacecraft had a separation of 44.5 degrees in the AB plane.}
\end{deluxetable}

\section{Observations}
The observations were conducted on 2008 January 10, as part of the Joint Observing Program (JOP) 200 - `Multi-point, High Cadence EUV Observations of the Dynamic Solar Corona'. Using the EUVI instrument \citep{wue04}, part of the Sun Earth Connection Coronal and Heliospheric Investigation \citep[SECCHI,][]{how08} instrument suite, to produce high cadence imaging of a sub-field region of interest. The aim of the STEREO/SECCHI led JOP was to use stereoscopic observations from the STEREO spacecraft, coupled to observations made in the Sun-Earth line, to produce a unique dataset examining the three-dimensional, time-dependent, nature of the solar atmosphere. The observations consist of a series of $2048\times512$ pixel, sun-centered, sub-field images in the $171$\mbox{\AA} passband, with a fixed exposure and cadence of 4~s and 30~s, respectively. Table~\ref{tab1} lists relevant observational parameters for both the STEREO A and B spacecraft. The high cadence observing program had a duration of two hours, beginning at 12:00:30UT for STEREO A and 12:00:50UT for STEREO B. The exposure timing is corrected for the light travel time between the two spacecraft, due to their differing distance from the Sun, ensuring that observations at both spacecraft locations originate from the Sun simultaneously.

Within these data, we analyze the active region loop system whose footpoints are centered on solar $x,y$ coordinates (-350\arcsec,10\arcsec) in STEREO A and (390\arcsec,10\arcsec) in STEREO B, where the coordinates are heliocentric-cartesian systems relative to the STEREO A \& B points of view. Figures~\ref{fig1} and \ref{fig2} show the relative location and geometry of the active region, as viewed from STEREO A \& B, with a spacecraft separation of 44.5 degrees in the A-B-Sun plane. This loop system shows a complex pattern of outward intensity propagations, similar to that observed by previous EUV imagers; for the first time, however, we observe these propagations simultaneously from two different points of view. A novel analysis technique is applied to investigate the three dimensional nature of these propagations as outlined in the following section. 

\section{Analysis}
\subsection{Preparation of the data}
The EUVI images must be cleaned and calibrated using a number of reduction procedures before the data can be analyzed. Using the SECCHI\_PREP routines available within the \emph{Solarsoft} database, the data are corrected for detector bias, flat field, and photometric calibration, which is applied to obtain the total number of counts per pixel. It is found that the series of images is affected by a systematic pointing `jitter', predominantly observed as a periodic variation of the pointing in the $x$ direction. This pointing jitter is corrected by applying a 2D cross correlation technique to long-lived, high contrast, features; in this case, stable plage structures observed at (335\arcsec,-10\arcsec) and (860\arcsec,20\arcsec) in the A and B images respectively. The $x$ pointing variation is found to have a sub-pixel amplitude on the order of a quarter of a pixel.

A 200$\times$200 pixel region of interest is selected, centered on the active region loop system in the A and B data. The solar rotation rate at the coordinates of the center of these sub-field regions is combined with the pointing jitter cross-correlation offsets and these combined offsets are used to co-align the images using cubic interpolation. Subsequently, two time series data cubes of 200$\times$200 pixels are then produced, centered on the active region loops, corrected and co-aligned as described above.

\begin{figure}[t]
\centering
\epsscale{1.1}
\plottwo{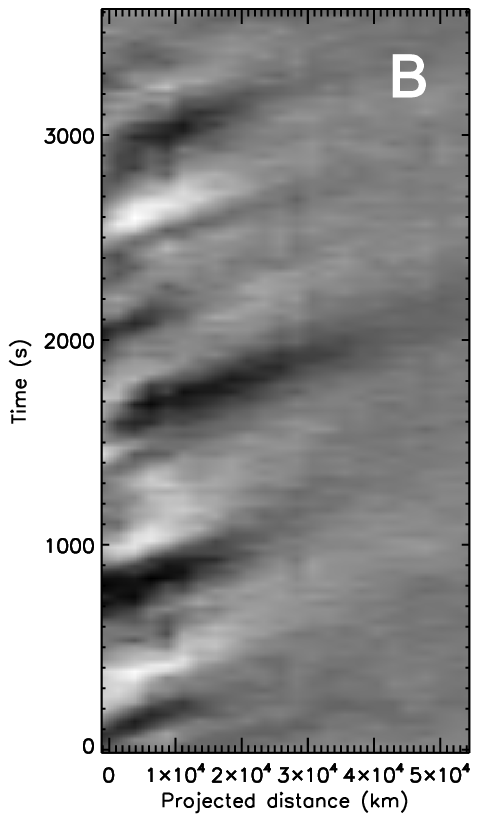}{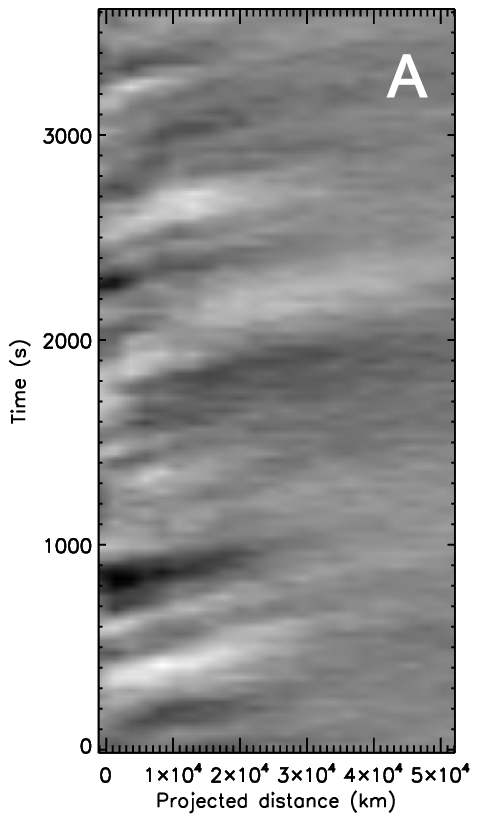}
\caption{Time-distance images formed from the integrated intensity along the loops outlined in Fig~\ref{fig2}, for STEREO B (left) and STEREO A (right). The abscissa indicates the observed distance along the loop, perpendicular to the line of sight from each spacecraft to the photosphere.}
\label{fig3}
\end{figure}

\subsection{Data analysis}
\subsubsection{Time-distance analysis}
To investigate the oscillatory nature of the propagations traveling along the coronal loop system, the loop intensity is integrated along the path of the loops. The loops are outlined by a pair of arcs that are parallel to the visible loop strands and enclose the observed propagations along them. These arcs are then joined by successive cross-sections, two pixels wide, progressing along the length of the arcs and perpendicular to the loop strands. The intensity of the pixels along each cross-section is integrated and divided by the number of cross-section pixels, to obtain the mean integrated loop intensity as a function of distance along the loop system. Figure~\ref{fig2} shows the location of the defined arcs enclosing the loops for both the A and B images, where the central axis is parallel to the observed propagations. These integrated loop intensity profiles are then used to produce time-distance images of the intensity propagations along the magnetic field lines.

\begin{figure}[t]
\epsscale{0.8}
\centering
\plotone{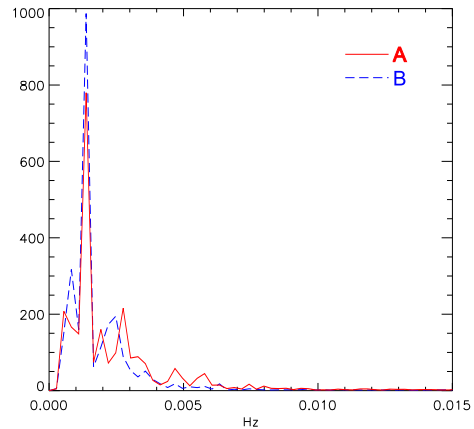}
\caption{Mean FFT of the integrated loop intensity from A (red) and B (blue), averaged over the lower $\approx$20,000~km as given in Fig~\ref{fig3}.}
\label{fig4}
\end{figure}

\begin{figure}[t]
\centering
\epsscale{1.1}
\plottwo{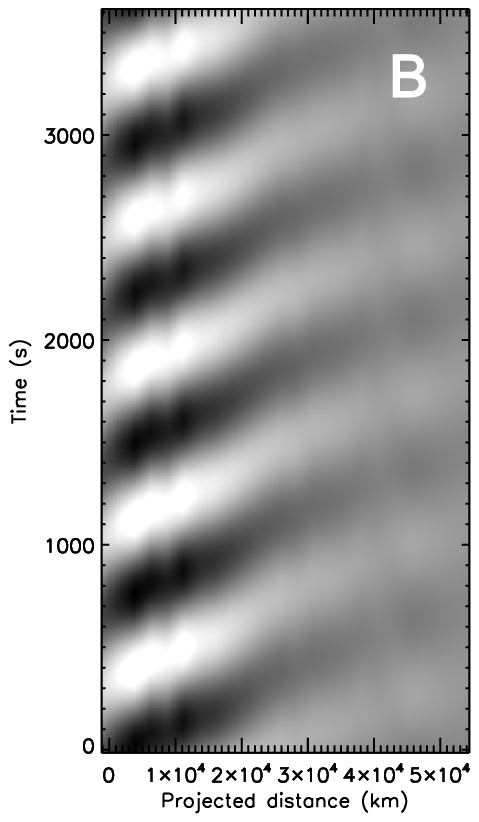}{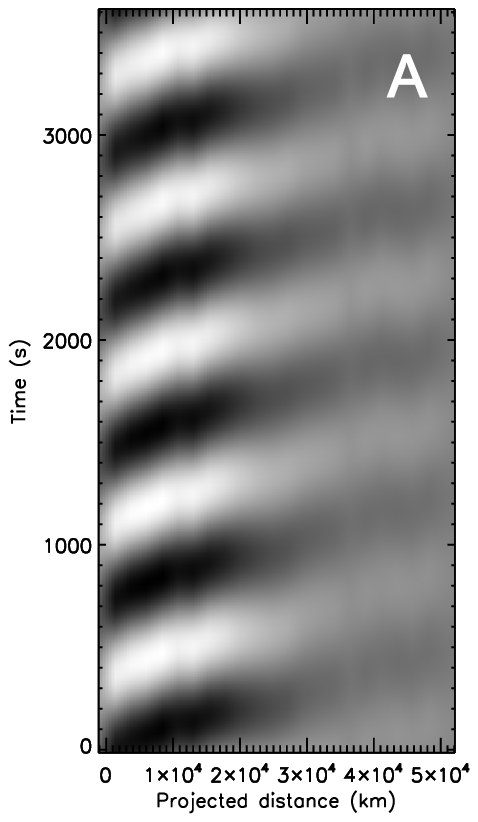}
\caption{Convolved time-distance images corresponding to Fig~\ref{fig3} filtered at 0.0014~Hz with a Gaussian filter of width 0.00014~Hz.}
\label{fig5}
\end{figure}

Figure~\ref{fig3} shows the time-distance images for the coronal loops outlined in Fig~\ref{fig2}, observed by the A and B spacecraft. Alternating light and dark bands of positive gradient indicate that a number of periodic outward intensity propagations are observed. They have a range of periodic timescales, most notably on the order of 700~s. Note the shallower gradient in STEREO A indicates a greater apparent velocity.

Figure~\ref{fig4} indicates the average Fourier transform of the integrated loop intensity profiles for A and B over the lower ten cross-sections of the time-distance images (or $\approx$20,000~km from the axes given in Fig~\ref{fig3}). It is clear to see that the dominant oscillatory power is observed at a frequency of 0.0014~Hz from both spacecraft. There appears to be power around the 5-minute range (2.4--4~mHz) as might be expected for this type of loop system \citep[cf.][]{dem02a}. However, the observations from each spacecraft show differing power distributions. This suggests that the structures carrying these higher frequency oscillations may be subject to line of sight effects from the point of view of each spacecraft, due to the differing temporal variability of the observed background plasma.

\subsubsection{Convolved time-distance analysis}
To investigate the spatial nature of the 0.0014~Hz oscillation in more detail, assuming stationary oscillations, a convolution filtering technique is applied to the time-distance images shown in Fig~\ref{fig3}. A Gaussian filter of 0.14~mHz width is multiplied by the Fourier transform of the time series at each spatial location of the time-distance image. The position of the Gaussian filter is progressively scanned through the frequency range to isolate the oscillatory power at a particular frequency. The resulting signal to noise of the oscillation is increased allowing greater precision estimates of the propagation velocity to be made. Figure~\ref{fig5} shows the time-distance images with the Gaussian filter centered on 0.0014~Hz, thus isolating the spatial structure of the dominant propagating oscillation indicated in Fig~\ref{fig4}.

\begin{figure}[t]
\centering
\includegraphics[width=16pc]{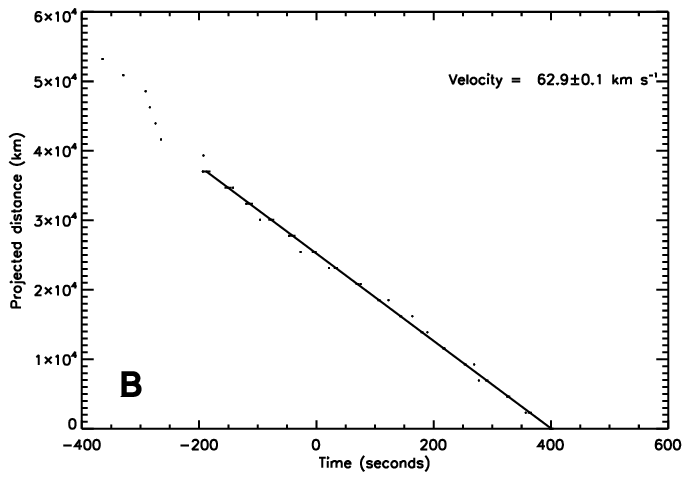}
\includegraphics[width=16pc]{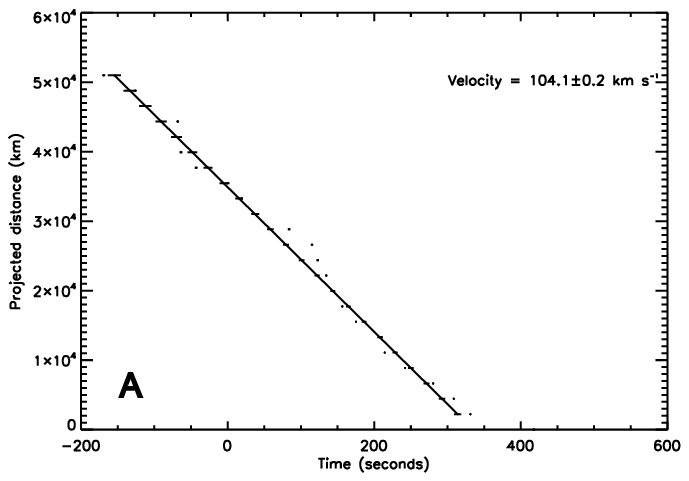}
\caption{Least squares fit of the projected distance along the loop system (perpendicular to the spacecraft line of sight) as a function of the propagation time from zero phase of the 0.0014~Hz oscillation, for STEREO B (top) and STEREO A (bottom).}
\label{fig6}
\end{figure}

\subsubsection{Propagation phase velocity}
To measure the apparent phase velocity of the 0.0014~Hz propagating oscillation observed from STEREO A and B, the oscillation phase is determined as a function of distance using the convolved time-distance images. The oscillation parameters at each spatial position are calculated by applying the Bayesian spectral analysis code described in \cite{me08}. A single frequency oscillation model is applied to the data to determine precise estimates of the oscillation parameters and their associated $1\sigma$ uncertainties. The oscillation phase propagation time is calculated from the oscillation phase, since the propagation time from zero phase is given by $\tau = \phi/\omega$, where $\phi$ and $\omega$ are the phase and angular frequency parameters determined by the Bayesian model. Regions where the phase jumps from 0 to $2\pi$ are corrected by subtracting $2\pi$, to provide continuity to the change in phase and phase propagation time. The phase velocity is then given by the inverse gradient from a least squares linear fit to the phase propagation time as a function of distance (Fig~\ref{fig6}).

\subsubsection{3D wave propagation}
Once the propagating phase velocity is known from both the STEREO A and STEREO B points of view, then the three dimensional velocity vector may be calculated stereoscopically. The stereoscopic 3D reconstruction software described in \cite{ash08b, ash08a} is used to determine the three dimensional vector along which the propagation travels \citep[see][for a detailed description of the reduction process]{ash08a}.

\begin{figure*}[t]
\epsscale{1.2}
\centering
\plotone{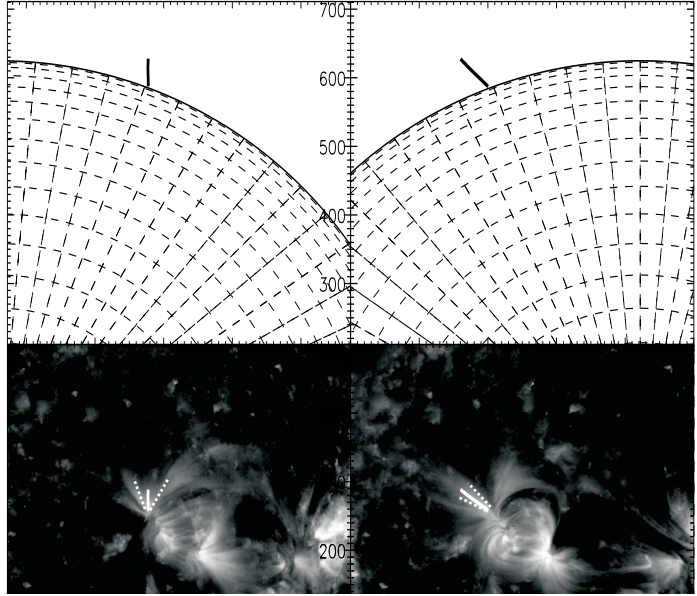}
\caption{Derived 3D geometry of the wave propagation along the coronal loops. Bottom panels show the observed region from STEREO A \& B overplotted with the projected propagation vector (solid line), and the arcs used to form the running difference analysis from Fig.~\ref{fig2} (dotted line). The top panels show the projected geometry of the propagation vector with the view rotated 90 degrees about the $x$-axis looking towards solar north.}
\label{fig7}
\end{figure*}

Following the propagation direction, vectors within STEREO A and STEREO B are defined using four point splines, these are parallel to the coronal loops, and are along the central axis of the regions used to form the time-distance images (shown in the bottom panels of Fig.~\ref{fig7}). These 2D vectors in the STEREO A and B image planes are combined stereoscopically to constrain the unique solution of the true propagation vector in three dimensions. The angle between the spacecraft line of sight and the 3D propagation vector, combined with the observed phase velocity, can be used to estimate the true propagation phase speed from each spacecraft.

\subsubsection{Error analysis}
The errors of the phase velocities measured in Fig.~\ref{fig6} are derived from the 1$\sigma$ errors on the filtered oscillation frequency and phase, returned by the Bayesian analysis code. These errors are propagated to determine the phase propagation time errors, which are applied in the least squares fitting routine used to determine the propagation phase velocity. As described in \cite{me08}, the Bayesian code is able to return precise estimates of the oscillation parameters. It is the precision of the phase and frequency estimates, combined with the increased signal to noise of the convolved time-distance analysis, that allows precise measurements of the propagation phase velocity.

The Aschwanden stereoscopy code determines an error on the three dimensional loop coordinates as the error along the line of sight in the reference frame of STEREO A. \cite{ash08a} estimates these height errors assuming that the position of a loop in the EUVI A \& B images can be selected to within one pixel by an automated loop fitting procedure. Due to the extended structure of the observed loop system, we estimate the loop vectors defined within A and B have a two pixel uncertainty. The error in the height of the points that define the three dimensional propagation vector are used to estimate the error range of the propagation inclination. The dependence on cosine results in an asymmetric error in the inclination. This error in the propagation inclination is propagated with the error in the measured phase velocity from A and B, to determine the error in the true propagation phase speed for each spacecraft.

\section{Results}
\subsection{0.0014~Hz wave propagation}
Figure~\ref{fig3} shows the time-distance images formed from the loops outlined in Fig.~\ref{fig2}. The complex pattern of periodic propagations indicates that a number of frequencies propagate upwards along the magnetic field of the loops. Figure~\ref{fig4} shows that the propagations are dominated by a frequency at 0.0014~Hz. The convolved time-distance images in Fig.~\ref{fig5} present the time-distance data isolated for the 0.0014~Hz frequency. It can been seen from Fig.~\ref{fig5} that the 0.0014~Hz oscillation propagates with a constant velocity along the loop when observed from both spacecraft, but that the propagation in the STEREO A observations has a shallower gradient than that observed from STEREO B. This suggests that the propagation has a greater apparent velocity perpendicular to the line of sight when viewed from STEREO A compared to STEREO B. Considering the observations of the loop system in Fig.~\ref{fig2}, this may be expected as the loop system appears to be more inclined away from the observers line of sight when viewed from STEREO A, thus increasing the velocity component perpendicular to the line of sight.

\subsection{Observed phase velocity}
Figure~\ref{fig6} shows the projected distance along the coronal loop system against the phase propagation time of the 0.0014~Hz oscillation. The gradient gives the phase propagation velocity in the plane perpendicular to the observation line of sight, where the negative value defines upward propagation from the base of the loop system. The data is fitted with a linear function (there is no significant acceleration trend as a function of distance along the loop), thus implying a constant phase velocity. As the time-distance images suggest, the phase velocity measured from STEREO A ($104.1\pm 0.2$ km~s$^{-1}$) is greater than that measured from STEREO B ($62.9 \pm 0.1$ km~s$^{-1}$), indicating that the propagation has a greater inclination to the line of sight when observed from STEREO A.

\subsection{3D wave propagation}
Using the stereoscopic reconstruction code \citep{ash08a}, the suggested propagation inclination relative to the A and B spacecraft is verified, with the propagation inclination of ${52.1^{+6.4}_{-5.4}} ^{\circ}$ relative to the STEREO A line of sight, an inclination of ${28.4^{+3.4}_{-2.1}} ^{\circ}$ relative to the line of sight of STEREO B, and an inclination of ${36.8^{+6.1}_{-5.1}} ^{\circ}$ to the local solar normal. The true propagation speed derived from the phase velocity observed from STEREO A is $132.0^{+9.9}_{-8.5}$ km~s$^{-1}$. Using the phase velocity observed from STEREO B, the derived true propagation speed is $132.2^{+13.8}_{-8.4}$ km~s$^{-1}$. Table~\ref{tab2} lists the observed velocities, relative geometry, and the absolute phase speed of the propagating wave.

\begin{deluxetable}{lcc}
\tabletypesize{\scriptsize}
\tablecaption{Parameters of the propagating wave\label{tab2}}
\tablewidth{0pt}
\tablehead{
\colhead{Propagation Parameter} & \colhead{STEREO A} & \colhead{STEREO B}
}
\startdata
Measured phase velocity (km~s$^{-1}$) & $104.1\pm 0.2$ & $62.9 \pm 0.1$ \\
Inclination to LOS (degrees) & ${52.1^{+6.4}_{-5.4}}$ & ${28.4^{+3.4}_{-2.1}}$ \\
Inclination to local normal (degrees) & ${36.8^{+6.1}_{-5.1}}$ & ${36.8^{+6.1}_{-5.1}} $ \\
Reconstructed phase speed (km~s$^{-1}$) & $132.0^{+9.9}_{-8.5}$ & $132.2^{+13.8}_{-8.4}$ \\
 \tableline
\enddata
\end{deluxetable}

\section{Discussion}
The results presented here represent the first measurements of the true propagation speed of the longitudinal slow mode within a coronal loop. The propagation speed of the slow mode can be derived from the dispersion relation for waves trapped within a magnetic cylinder \citep{edw83}. The slow mode propagates within a cylindrical geometry with a phase speed $c_{t} \le v_{ph} \le c_{s}$. In the long wavelength limit, in comparison with the diameter of the structure, the slow mode propagates at the tube speed $c_{t}$, and in the short wavelength limit propagates at the sound speed $c_{s}$. The waves observed here tend to the long wavelength limit and, strictly, propagate at the tube speed 
\begin{equation}\label{eqn1}
c_{t}^{2}=\frac{c_{s}^{2}\nu_{a}^{2}}{c_{s}^{2}+\nu_{a}^{2}},
\end{equation}
 where $\nu_{a}$ is the Alf\'{v}en speed. In the case where $v_{a} \gg c_{s}$, then $c_{t} \equiv c_{s}$.

If we assume a low-beta coronal plasma with a typical Alf\'{v}en speed of 1000~km~s$^{-1}$ and the measured tube speed of 132~km~s$^{-1}$, the tube and sound speed differ by less than 1$\%$; thus we may approximate the tube speed to be equal to the sound speed. The sound speed for a weakly bound collisional plasma such as the low corona may be approximated using the ideal gas equation of state, to derive the temperature.
\begin{equation}
c_{s}=\sqrt{\frac{\gamma p}{\rho}}=\sqrt{\frac{\gamma N_{\mu} k T}{\mu m_{p}}},
\end{equation}
\begin{equation}\label{eqn3}
T=\frac{c_{s}^{2} \mu m_{p}}{\gamma N_{\mu} k},
\end{equation}
 where the adiabatic index $\gamma=5/3$, $p$ is the plasma pressure, $\rho$ is the plasma density, $k$ is the Boltzmann constant, $T$ is the temperature, $m_{p}$ is the proton mass, and assuming a fully ionized hydrogen and helium plasma of coronal abundances, the mean molecular weight $\mu=1.27$ and the mean particle number $N_{\mu}=1.92$. 
 
\begin{figure}[t]
\epsscale{1.0}
\centering
\plotone{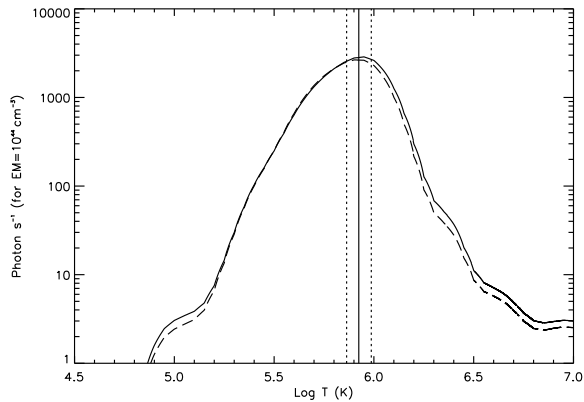}
\caption{Response functions of the EUVI $171$\mbox{\AA} channel for STEREO A (solid) and STEREO B (dashed). The vertical lines indicate the loop temperature (solid) and errors (dotted), inferred from the STEREO A phase speed results.}
\label{fig8}
\end{figure}

The measured phase speeds of $132.0^{+9.9}_{-8.5}$ km~s$^{-1}$ and $132.2^{+13.8}_{-8.4}$ km~s$^{-1}$, from STEREO A and B respectively, are used to derive infer the temperature of $0.84^{+0.13}_{-0.11}$~MK and $0.84^{+0.18}_{-0.11}$~MK. This derived temperature is very close to the peak of the EUVI $171$\mbox{\AA} response functions, as shown in Fig.~\ref{fig8}. The $171$\mbox{\AA} bandpass is dominated by emission from the \ion{Fe}{9} $171$\mbox{\AA} line formed around these temperatures, suggesting that the waves are observed by emission from \ion{Fe}{9} $171$\mbox{\AA}. The correspondence between the sound speed at the peak response temperature and the measured phase speed gives further evidence that the slow magnetoacoustic mode is observed.

It should be remembered, however, that these results are based on the assumption of low plasma beta. It is conceivable that over-dense structures may exist in the corona, in which case the Alf\'{v}en speed can be of the same order of magnitude as the sound speed. The tube and sound speeds may then be significantly different, from Equation~\ref{eqn1}. With accurate spectroscopic estimates of the temperature and density, this allows the possibility of discriminating between the tube and sound speeds for seismological purposes, such as determining the magnetic field strength, or adiabatic index, for example; this will be addressed in a following paper.

The large variation in the observed phase velocities from STEREO A \& B demonstrates the significance of the projection effect on previous observational studies of longitudinal modes using EIT and TRACE, such as \cite{dem02a}. The projected phase velocity in the plane of the sky cannot be approximated to the phase speed and used to derive meaningful physical parameters. Using single point observations, it is not possible to disentangle the projection effect to study the physical causes of any variation in observed phase speeds.

The results for the phase speed derived from both the STEREO A and STEREO B observations agree very well within the errors. This close agreement between the results suggests that the correct 3D propagation geometry has been determined and validates the implementation of the \cite{ash08a} stereoscopic code.

The constant velocity shown in Figs.~\ref{fig5} and \ref{fig6}, combined with the fact that the propagation follows a linear path from the stereoscopic results, suggests that a constant adiabatic index and equilibrium temperature is maintained along the loop strands, probably due to the high thermal conduction along the magnetic field.

There has been a limited number of previous observations of low frequency oscillations of this order in coronal loops. For example, \cite{terr04} find an $\simeq$0.0016~Hz oscillation in TRACE observations of fan like loops, \cite{mci08} also observe a 0.0015~Hz oscillation in TRACE loops associated with plage. It is possible that larger scale studies such as that of \cite{dem02a} could fail to detect these low frequency oscillations due to a selection effect of using a wavelet analysis and time series of limited duration, since wavelet analysis has a detection cutoff at low frequencies, due to the cone of influence of the wavelet transform. Also, the use of running difference analysis over a time range designed to detect periodicities in the range 3-5 minutes will selectively exclude other periods and detection of lower frequencies.

\section{Conclusions}
The measurement of the phase speed of the longitudinal slow magnetoacoustic mode in coronal loops is presented for the first time. A 0.0014~Hz intensity oscillation propagating within a coronal loop system is observed by both EUVI instruments on the STEREO A and B spacecraft. High precision estimates of the wave phase velocity, observed in the plane of the sky from each spacecraft, are determined using convolved time-distance images and a Bayesian oscillation code. The absolute phase speed is then calculated using the three dimensional geometry of the propagation obtained from the stereoscopic observations.

The measured phase speeds of $132.0^{+9.9}_{-8.5}$ km~s$^{-1}$ and $132.2^{+13.8}_{-8.4}$ km~s$^{-1}$ are consistent with the expected speed of the slow magnetoacoustic mode. The inferred loop strand temperature of $0.84^{+0.13}_{-0.11}$~MK and $0.84^{+0.18}_{-0.11}$~MK, under the assumption of low plasma beta, is close to the peak of the $171$\mbox{\AA} response function. Thus, the observed phase speed corresponds to the sound speed at this temperature, further strengthening the slow magnetoacoustic mode interpretation. 

In principle, it is possible to use the propagating slow mode to measure the coronal magnetic field strength. Temperature and density diagnostic observations are necessary to determine if the slow mode tube speed can be discriminated from the sound speed within the errors. This possibility is currently under investigation and will be reported in a following paper. 

\acknowledgments
This research is supported by the Science and Technology Facilities Council (STFC) under grant number ST/F002769/1.
The STEREO/SECCHI data used here are produced by an international consortium of the Naval Research Laboratory (USA), Lockheed Martin Solar and Astrophysics Lab (USA), NASA Goddard Space Flight Center (USA) Rutherford Appleton Laboratory (UK), University of Birmingham (UK), Max-Planck-Institut f\"{u}r Sonnensystemforschung(Germany), Centre Spatiale de Liege (Belgium), Institut d'Optique Théorique et Applique\'{e} (France), Institut d'Astrophysique Spatiale (France).
Tutorials for the Ashwanden stereoscopic reconstruction software are available at
\anchor{http://www.lmsal.com/~aschwand/stereo/stereo_soft/software2.html}{http://www.lmsal.com/\\$\sim$aschwand/stereo/stereo\_soft/software2.html}. 
M.S. Marsh would like to acknowledge useful discussion with V.M. Nakariakov and the encouragement of L.E. Pickard.

{\it Facilities:} \facility{STEREO (EUVI)}.

\bibliographystyle{apj}
\bibliography{ms}

\end{document}